Time-lapse Study of Neural Networks Using

Phase Imaging with Computational Specificity (PICS)

By

Eunjae Kim

Senior Thesis in Computer Engineering

University of Illinois at Urbana-Champaign

Advisor: Gabriel Popescu

May 2020

# Abstract

In life sciences, fluorescent labeling techniques are used to study molecular structures and interactions of cells. However, this type of cell imaging has its own limitations, one of which is that the process of staining the cells could be toxic to the cells and possibly damage them. We are specifically interested in time-lapse imaging of live neurons to study their growth and proliferation. Neurodegenerative diseases are characterized by phenotypic differences in neuron growth and arborization.

This thesis proposes a label-free digital staining method using the deep convolutional neural network to address the issues with the previous cell imaging method. Our results show that a deep neural network, when trained on phase images with correct fluorescent labels, can correctly learn the necessary morphological information to successfully predict MAP2 and Tau labels. This inference, in turn, allows us to classify axons from dendrites in live, unlabeled neurons.

Subject Keywords: Quantitative phase imaging; Gradient light interference microscopy; Time-lapse microscopy; Neuronal growth analysis; Machine Learning; Semantic segmentation

# Acknowledgments

I would like to express my sincere gratitude to my advisor, Professor Gabriel Popescu, for his guidance through each stage of the process over the past academic year. I would also like to thank two graduate students at the QLI Laboratory, Mikhail Kandel and Youngjae Lee, for their intellectual support and critical feedback.

# Contents

# 1. Introduction

As in many other biological studies, the study of neurons often involves immunohistochemical staining and fluorescent microscopy, and these methods have been proven very effective. However, photobleaching limits the duration of time over which fluorescence imaging can be used, and the neuron cultures can be damaged during or after the staining process due to phototoxicity [1-3]. Quantitative phase imaging (QPI) is a label-free imaging method that provides high contrast images in a non-destructive manner [4, 5], which makes it a good method for time-lapse imaging [6-12]. However, QPI lacks the specificity that chemical staining can provide. We propose that computational methods can be used to augment quantitative phase imaging and provide specificity by computationally generating the staining labels. In this research, we conduct the time-lapse study of hippocampal neurons using phase imaging with computational specificity, a recently developed microscopy method [13].

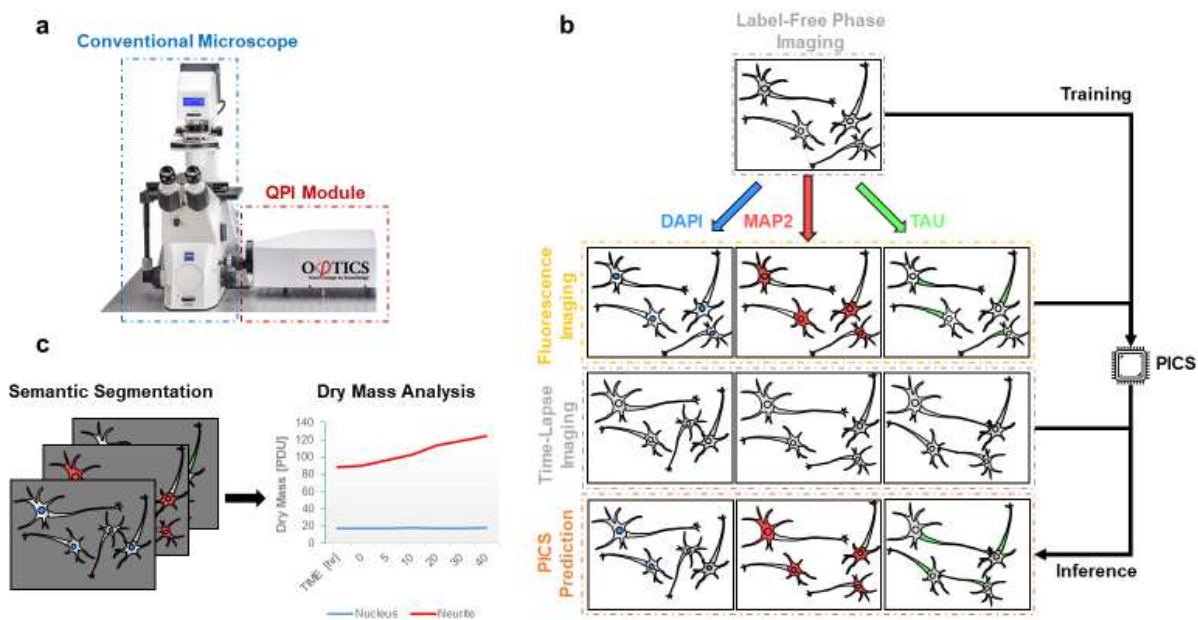

Figure 1. Overview of the PICS method used for the neural network study. (a) The conventional transmitted light microscope with the QPI module. Used with the permission of Ref. [13]. (b) The process of computationally generating fluorescent labels on time-lapse data. (c) Applications of the method to the semantic segmentation and dry mass analysis.

We will specifically study the growth and arborization of the neuronal sub-cellular compartments. We will first observe the neuron culture over a long period of time using time-lapse, label-free phase imaging. Afterward, the neurons will be stained with Tau and MAP2 proteins, which tag axons and dendrites in the neurons [14]. These stains will produce the labels for those compartments when imaged with a fluorescence microscope by exciting the fluorophores with light. Then, we will use a deep convolutional neural network to learn the relationship between the texture and morphology in the phase image and the fluorescence labels. The deep convolutional neural network will be able to generate the fluorescent labels from phase images, and we can use these labels to perform semantic segmentation and growth analysis on the neuron culture. Figure 1 shows the general procedure of the method we use.

# 2. Literature Review

## 2.1 Tau and MAP2 Proteins

Microtubules are components of the cytoskeleton in neurons that mediate many important functions including organelle transport, neurite elongation, cellular structure, and maintenance as well as neurite elongation [15]. The microtubules can be modulated by interactions with microtubule-associated proteins (MAPs).

Among many MAPs, Tau and MAP2 are now studied more extensively than the others due to their ability to bind and stabilize and increase the rigidity of microtubule [16]. Tau and MAP2 have different microtubule-binding repeats near the carboxyl terminus that make neurite differentiation to axon and dendrites [17]. In cortical and hippocampal neuronal cultures, Tau proteins segregate into axons, while MAP2 stays into the nascent dendrite [18]. Thus, in the mammalian nervous system, Tau proteins are expressed in all neurons including the cell body, dendrite and axon while MAP2 is expressed in dendrites and the cell body.

It is important to study these two proteins to understand and distinguish the location of these two proteins through neural networks, especially synaptic plasticity. Tau regulates microtubule dynamics including outgrowth of axons and axonal transport that is related to the stability of axonal microtubules while MAP2 mediates dendrite outgrowth and located mainly in dendritic microtubule [19].

Dysfunction in microtubules impairs axonal transportation of vesicles and organelles, which can lead to apoptosis of neurons [19]. It has been reported that dysfunction of Tau protein leads to neurodegeneration such as Alzheimer's disease, frontal temporal dementia and Pick's disease [17, 19, 20]. Many post-mortem studies of brains showed that dysfunction of Tau and MAP2 reduces synaptic function and neuronal connectivity that leads to the loss of neurons [21, 22].

## 2.2 Fluorescence Microscopy

Fluorescence microscopy is an important tool for modern biological studies. Fluorescence microscopy enables researchers to comprehend cell physiology and proliferation by providing direct visualization of the subcellular structures and activities [23]. Fluorescent microscopy uses a wide range of fluorescent indicators to target certain proteins, lipids or ions in specimens and illuminate only the area of interest. Fluorescent microscopy is an outstanding modality of biological imaging because of the high specificity it offers.

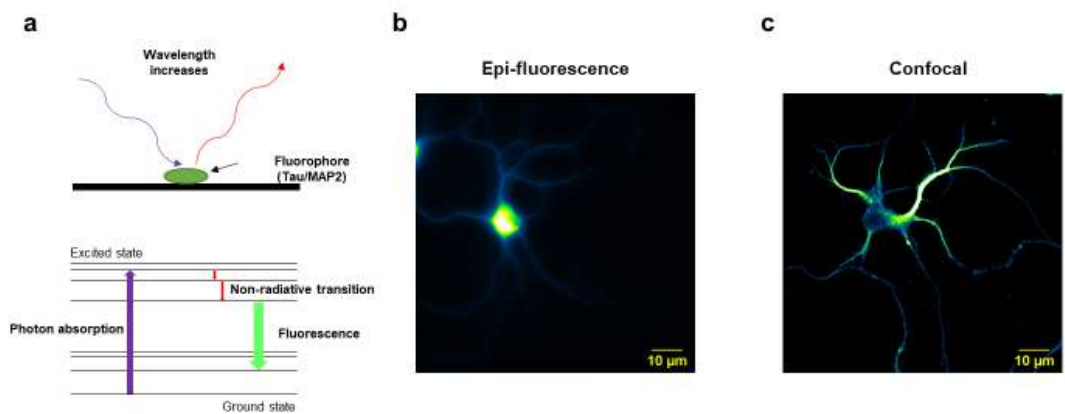

Figure 2. Overview of fluorescence microscopy. (a) Fluorescence and Jablonski diagrams. (b) An image of a neuron taken with an epi-fluorescence microscope. (c) An image of a neuron taken with a confocal microscope.

For single-photon microscopy, in principle, fluorophores (indicators) absorb light and emit light of longer wavelengths, where the wavelengths of light get longer because of the lost energy in the absorption-emission process (Figure 2(a)). This change in wavelengths is called the Stokes shift. To obtain fluorescence images of the specimens, the excitation light needs to be separated from the emission light. Optical filters and dichroic mirrors are used for this separation in many fluorescence microscopes [24].

In the case of two-photon, or multiphoton, microscopy, a fluorophore absorbs multiple photons at the same time and emits a photon with larger energy than the individual photons absorbed. The advantage of such a system is that it can use light with a longer wavelength as an excitation source and light with a longer wavelength can penetrate deeper into specimens. Also, it causes less damage to the specimens as it has less energy [24].

Epifluorescence microscopy, also referred to as widefield microscopy, is a commonly used mode of fluorescence imaging. The excitation light from the fluorophore in the specimens is focused on the detector through the same objectives as the illumination light [25]. It is a quick, simple method to capture fluorescent images (Figure 2(b)). However, it is often subject to high background signals caused by the illumination and excitation outside the focal plane.

Confocal microscopy aims to reduce out-of-focus light and improve image quality. Confocal microscopy uses a laser light source and uses a pinhole aperture to ensure that only the excitation light of the correct focus reaches the detector [26]. Images are acquired by acquiring the intensity at each point sequentially. Confocal microscopy can achieve high spatial resolution and contrast in fluorescent imaging (Figure 2(c)), but the acquisition process is much more complicated.

Stimulated emission depletion (STED) microscopy is another technique in fluorescence microscopy for super-resolution microscopy. The resolution gain is achieved by only activating a very small area of the sample. The doughnut-shaped STED beam realized with phase masks depletes the fluorescence signal and reduces the excitation area [27].

A fluorophore is a chemical compound that is an essential part of fluorescence microscopy. A fluorophore absorbs light and emits another light with a different wavelength. Each fluorophore has specific wavelengths of light that it takes for excitation and emission. Different fluorescence microscopy methods require different types of fluorophores [28].

Fluorophores are generally divided into the following groups: biological fluorophores, organic dyes, and quantum dots. Biological fluorophores are fluorescent proteins such as green fluorescent protein (GFP) from the jellyfish Aequorea Victoria [29]. The fluorescent proteins bind to biological molecules of interest and allow researchers to observe biological events. Organic dyes are synthetic fluorescent compounds, such as fluorescein. These small fluorophores can be linked to macromolecules and used for observation of areas of interest without biological interference. Quantum dots are nanoscale semiconductors. The size of a quantum dot determines the excitation and emission wavelengths [30, 31].

## 2.3 Quantitative Phase Imaging (QPI)

While fluorescence microscopy is an essential method to study subcellular components with high specificity, there are some intrinsic limitations of the method [1-3]. Photobleaching is the photochemical alteration of the fluorescence indicator which causes the indicator to lose the ability to fluoresce. Phototoxicity refers to the cellular and subcellular damage caused by the exposure of the specimen to the high-intensity illumination. Quantitative phase imaging can be used to avoid such problems associated with fluorescence microscopy.

Quantitative phase imaging (QPI) is an emerging label-free microscopy method. It is a powerful method to obtain high contrast images of thin, transparent specimens [4, 5]. Based on Abbe's theory of image formation as an interference phenomenon, Zernike developed phase contrast microscopy (PCM) [32].

PCM achieves high image contrast using phase shifts [5]. Later, Gabor proposed a method of holography that allows one to store phase information quantitatively, though it was originally intended for a different use [33]. QPI extended these ideas to develop a high contrast imaging method using quantitative information such as thickness and refractive index of the specimens.

Over the past decades, many approaches to QPI measurements have been demonstrated. Diffraction phase microscopy (DPM) is a common-path, off-axis QPI method that significantly alleviates the phase noise problem (Figure 3) [34]. The configuration of DPM cancels out most of the mechanisms that cause the noise. DPM is built on top of a bright-field microscope. Through the transmission grating at the output of the bright-field microscope, only the selected diffraction orders are passed. A DC reference field (unscattered) and a sample field (scattered) are created at the Fourier plane using a physical pinhole or a spatial light modulator (SLM) [35]. These different components are combined into an interferogram and used to reconstruct a phase map later. The method to perform the reconstruction in real-time has been previously demonstrated [36].

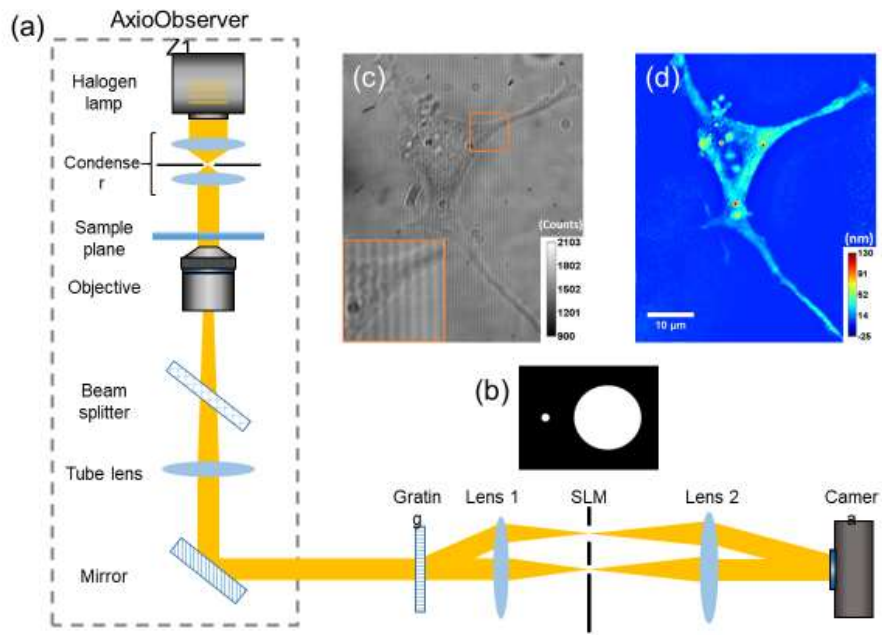

Figure 3. Schematic of DPM. (a) DPM is off-axis, common-path module attached to a conventional bright-field microscope. (b) The sample and reference field are created by spatial filtering at the Fourier plane. (c) A raw interferogram and (d) its corresponding optical pathlength map. Used with permission from Ref. [37].

Spatial light interference microscopy (SLIM) is a typical common-path, phase-shifting QPI method (Figure 4) [38, 39]. SLIM is implemented as an add-on module to an existing phase contrast microscope, and as a speckle-free imaging method with exceptionally low spatial background noise. SLIM combines the classic ideas of Zernike's PCM and Gabor's holography. SLIM also utilizes the spatial decomposition of the image field into two sample and reference components. At the output of a phase contrast microscope, Fourier lenses are used to introduce the phase shifts. Here, an additional phase shift is provided by a liquid crystal phase modulator (LCPM). Four intensity images corresponding to each phase delay are acquired as a result. Many successful use cases of the phase images from SLIM for morphological analysis and growth dynamics measurements have been reported [12, 40-42].

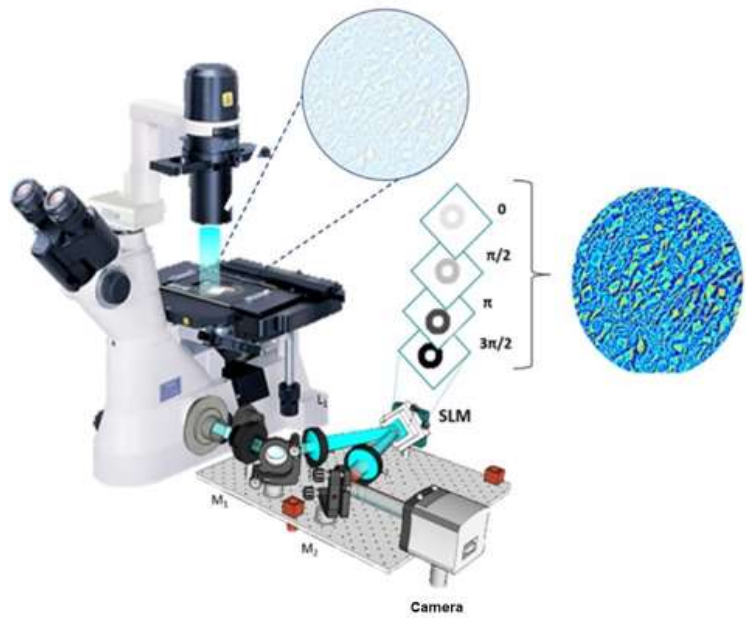

**Figure 4. System schematic of SLIM. Used with permission from Ref. [45].**

Gradient light interference microscopy (GLIM) combines diffraction interference contrast (DIC) microscopy with low-coherence interferometry and holography to solve the problem with multiple scattering that affects image contrast (Figure 5) [43]. GLIM is also designed as an add-on module and is suitable for high contrast imaging of thick specimens. GLIM also generates four images with different phase shifts applied by the SLM, and a phase-gradient map is obtained from those images. The phase-gradient map is later integrated to recover the quantitative phase map. GLIM has been proven to be a perfect solution to image thick and opaque samples with high sectioning and resolution capabilities [7, 44].

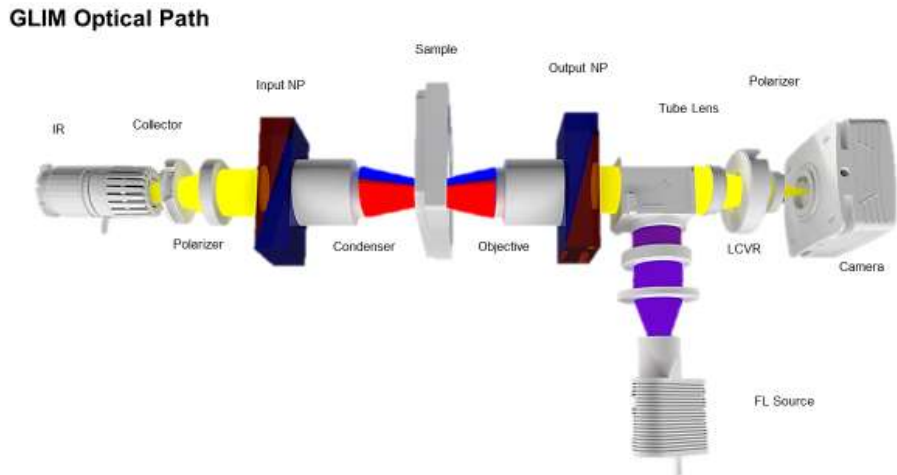

Figure 5. System schematic of GLIM. Used with permission from Ref. [13].

As the field of QPI grows rapidly, it has been shown that QPI can be a suitable method to study biological samples. QPI, as a high-contrast label-free imaging method, can be used to study cellular structures and dynamics of live, unstained cells. With the right analysis method, the quantitative information extracted from the phase map can provide insightful knowledge. Many successful use cases have been demonstrated in the studies, such as cellular dynamics, cell growth, and proliferation, mass transport, cell-substrate interaction, the impact of the extracellular environment, pathology, etc. [45-52].

## 2.4 Phase Imaging with Computational Specificity (PICS)

In parallel with the development of QPI, the field of artificial intelligence has been expanding quickly with fruitful applications to many other areas. There have also been many efforts to combine AI techniques with QPI [53-55]. Built upon those previous ideas, the research group of the author has recently developed a new method to combine quantitative phase imaging and deep learning, called phase imaging with computational specificity (PICS).

PICS aims to take advantage of both popular imaging methods, fluorescence microscopy and quantitative phase imaging. They each have clear advantages and disadvantages. Fluorescence microscopy provides high specificity, but it is not suitable for time-lapse imaging because of photobleaching and phototoxicity. Quantitative phase imaging can extract quantitative information from the specimen with high accuracy, but it lacks specificity. PICS offers a solution that is based on the label-free phase imaging with the specificity introduced by digitally generated fluorescent stains using deep learning.

# 3. Research Materials and Methods

## 3.1 Neuron Culture and Immunohistochemistry

For neuron imaging, the hippocampal neurons were prepared as follows. Primary hippocampal neurons were harvested from dissected hippocampi of Sprague-Dawley rat embryos. Hippocampi were dissociated with enzymes in order to have hippocampal neurons. Hippocampal neurons were then plated on to 6 well plates that are pre-coated with poly-D-lysine (0.1 mg/ml; Sigma-Aldrich). Hippocampal neurons were initially incubated with a plating medium containing 86.55% MEM Eagle's with Earle's BSS (Lonza), 10% Fetal Bovine Serum (re-filtered, heat-inactivated; ThermoFisher), 0.45% of 20% (wt./vol.) glucose, 1x 100 mM sodium pyruvate (100x; Sigma-Aldrich), 1x 200 mM glutamine (100x; Sigma-Aldrich), and 1x Penicillin/ Streptomycin (100x; Sigma-Aldrich) in order to help attachment of neurons (300 cells/mm2). After three hours of incubation in an incubator (37°C and 5% $CO_2$), the plating media was aspirated and replaced with maintenance media containing Neurobasal growth medium supplemented with B-27 (Invitrogen), 1% 200 mM glutamine (Invitrogen) and 1% penicillin/streptomycin (Invitrogen) at 37 °C, in the presence of 5% $CO_2$.

In this study, the fluorescent Tau antibody (ab80579) and MAP2 antibody(ab32454) were used to stain axon and dendrite accordingly. The application of the antibody was adapted and modified from established protocol from the ThermoFisher and Abcam website. Hippocampal neurons were grown for 4 days and we started to image GLIM time-lapse images for 4 more days in an environment of 37 °C with 5% $CO_2$.

Neurons were then fixed with freshly prepared 4% paraformaldehyde (PFA) for 15 minutes following 0.5% Triton-X for 10 minutes and 2% BSA for 30 minutes incubation. Hippocampal neurons were incubated in the anti-Tau antibody that was diluted to 1:250 in 1% BSA for 8 hours following Goat anti-mouse secondary antibody (ab205719) incubation that was diluted to 1:500 in 1% BSA for 8 hours.

Hippocampal neurons were then incubated in the anti-MAP2 antibody that was diluted to 1:500 in 1% BSA for 8 hours following Goat anti-rabbit secondary antibody(ab205718) incubation that was diluted to 1:1000 in 1% BSA for 8 hours.

## 3.2 Data Acquisition and Preprocessing

The four phase images acquired from the GLIM system were used to reconstruct the quantitative phase map and later integrated with the Hilbert transformation to form the final phase image used in this research, as shown in Figure 6(a). The fluorescent images contain two channels for Tau and MAP2 labels. These pairs of phase images and fluorescent images can be used to train the deep convolutional neural network after some preprocessing steps along the process.

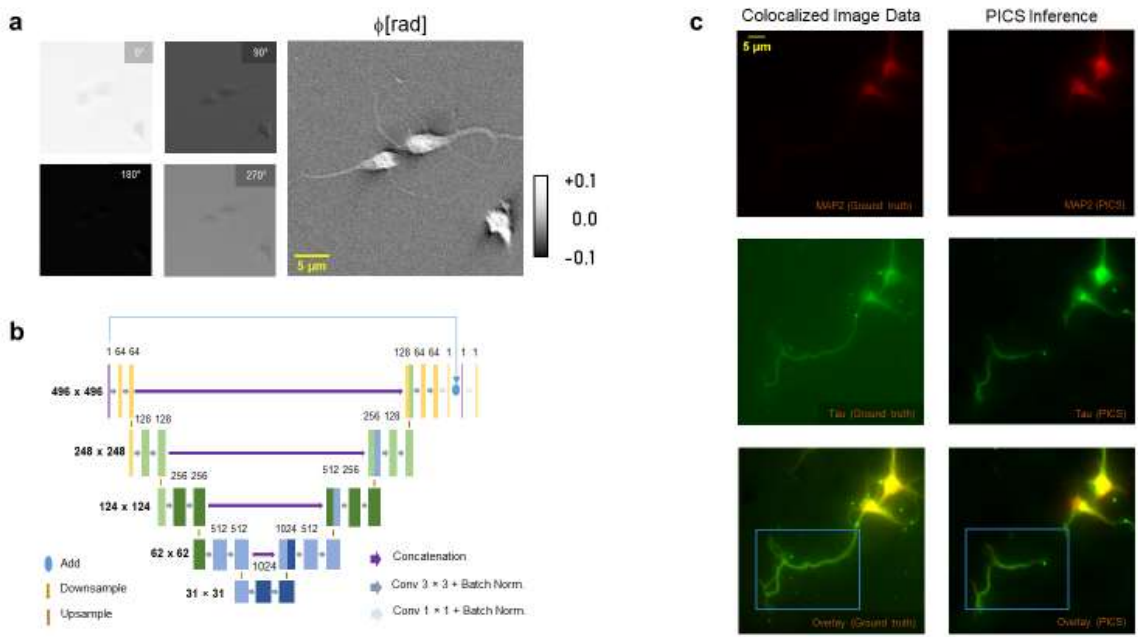

Figure 6. Training overview. (a) Four phase images from the GLIM system and reconstructed phase image. (b) The network architecture based on U-Net. (c) The ground truth data (left) and the prediction results (right).

During the imaging process, some level of background noise is inevitable. The background noise might come from many sources including the dust and water drop formed on the surface of the glass that contains the sample. Background subtraction was performed to get rid of such noise that is common across the dataset. The background image was calculated by taking the average of the images excluding the outliers in terms of the energy of the image, which is defined by the sum of the absolute value of each pixel.

Each microscope field was captured across multiple focus levels, producing multiple images of the same field with a different focus (z-stacks). There has been an approach that uses the z-stacks of images to train the deep convolutional network [56]. Though it has its own advantages, the approach will require a larger size of the model to take the full advantage. In this research, we selected the single focus level that is used for both training and inference. The focus level was selected using the 2-D discrete wavelet transform with Haar wavelet filters. The phase images and fluorescent images are colocalized using intensity-based image registration to resolve the unknown level of shift introduced to the fluorescent channels. Then, images were cropped by 64 pixels, from 2048x2048 to 1984x1984, to address the shift after registration.

### 3.3 Network Architecture

These colocalized pairs of phase images and fluorescent images are used to train deep convolutional neural networks for supervised regression machine learning. The phase images are fed to the network and the corresponding fluorescent labels are also seen by the network. The deep convolutional neural network will be able to learn, if any, the predictive relationships between the quantitative information in the phase image and the fluorescent labels.

In this research, we used U-Net for the main architecture for the deep convolutional neural net. U-Net was originally developed for the medical image semantic segmentation solution and is characterized by the skip connections added between the decoding and encoding paths of the network [57]. U-Net is also widely used as a backbone network structure for image-to-image translation problems [58].

We used a modified version of U-Net that was introduced in the original PICS paper [13] with batch normalization added in each layer and a direct add connection between input and output layers. The detailed network structure is shown in Figure 6(b). We trained two such networks separately each of which was trained with phase-Tau pairs and phase-MAP2 pairs. For this work, we explored multiple variations of ways to train these networks, which will be discussed further in the next section. We could also train a single network with a slight modification for the output layer where the network would output an image with two channels instead of one. However, our results showed that this does not improve the performance and rather increases the GPU memory constraint.

The trained deep convolutional network models can be used to make an inference on other unseen phase images to predict the Tau and MAP2 labels. By the nature of the convolutional neural network, the network can make inference on images of any size if the input images share the same characteristics as the phase images it was trained on. We can make such inference the time-lapse phase images of live unstained biological neural networks taken over many cell cycles to study neuronal growth.

## 3.4 Training

We obtained 243 pairs of 1984x1984 (pixels) images of hippocampal neurons. 218 of them were used for training, in which 10% was reserved for validation, and 25 of them were used for testing. The left column of Figure 6(c) shows the example fluorescent images for Tau and MAP2 labels. It shows that the Tau label is larger and includes most of the MAP2 label since it stains both axons and dendrites. Our goal is to train our networks so that it can learn the morphological information associated with each label from the phase image. The right column of Figure 6(c) shows a successful inference from our trained convolutional neural network on the image that was unseen to the model. There is a strong overlap of the label on the cell body while there are very little overlap and strong Tau signal along the axon. It signifies that our network can distinguish the characteristics of the sub-cellular structures from the quantitative information captured in the phase image.

To train the neural network model, we cut the images in the training data into four smaller patches to reduce GPU memory usage for each step without loss of performance. We obtained 872 of 992x992 images for training, where again 10% was reserved for validation. Also, each image was downsampled by the factor of 2 for the same reason. The varying patch size can affect the performance of the network. Our experiments showed that using smaller patches can improve the performance, but the patch size still must be large enough to convey the necessary morphological information of the neuronal sub-compartments. In fact, our network in the given settings had the best performance using the 992x992 patches downsampled to 496x496 images.

In addition to the network structure and the input image patch size, we explored several options to improve the performance of our network. We mainly used traditional L1 loss for optimization since it has been shown that using L1 loss results in less blurring compared to L2 loss [58] for regression problems like this. We tried varying the size of the model, i.e., the number of the parameters, since it has been shown that the reduced model size can prevent overfitting without loss of performance [13]. However, our result shows that using the larger model has a better performance. It indicates that our model needs to be big enough capacity to learn the subtle difference in the texture and morphology, which is required to identify the neuronal structures of interest in the phase images and correctly label them. We also tried combining the L1 loss with other loss functions. We compared the performance of the network when trained with the L1 loss only, the combination of L1 loss and Pearson correlation coefficient loss, and the combination of L1 loss and GAN loss [58]. To take advantage of the GAN loss, we adapted out network model into the generator of a conditional GAN model where 64x64 PatchGAN was used for the discriminator. The predictions of MAP2 label using the different methods described above are shown in Figure 7(a), and Figure 7(b) is the visualization of each layer for an example image. In our experiments, we found that using the weighted sum of the L1 loss and Pearson correlation loss gives the best performance in terms of the correlation between the predicted labels and the ground truth labels.

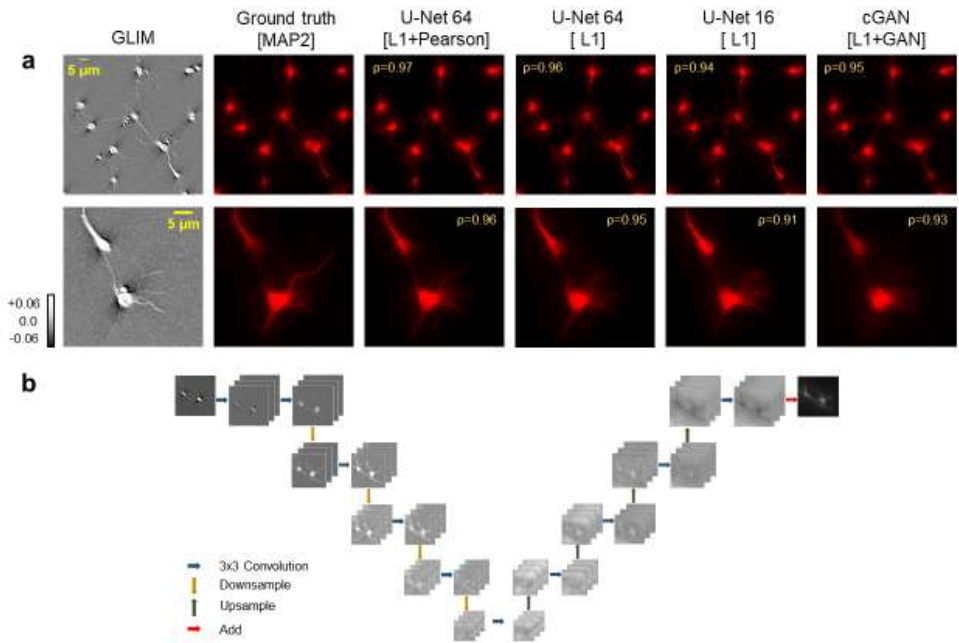

Figure 7. (a) A comparison of the results from different models and loss functions explored in this research. (b) A visualization of intermediate features from each layer.

# 4. Research Results

## 4.1 Time-lapse Inference

The deep convolutional neural network model trained on the stained ground truth neuron samples can be used to make inference on the time-lapse imaging data of live, unstained neurons. The time-lapse data consist of phase images of live neurons taken over a long period of time without any detrimental effects of photobleaching or phototoxicity typically accompanied by fluorescence imaging. We have obtained phase images of neurons in vitro for 40+ hours.

The inference result on the time-lapse data is shown in Figure 8. The red and green labels in the middle rows correspond to the MAP2 and Tau labels computationally generated by our method. The main difference between the two labels, as discussed earlier, is that the Tau label generally includes MAP2 label and Tau additionally stains axons. Thus, in the overlay images of the generated labels at the bottom row, the non-overlapping region with a strong green signal must correspond to the axon. On the other hand, the yellow region of the overlay image is the part of the neuron stained by both Tau and MAP2, which indicates that the specific compartment belongs to the cell body or dendrites. In the right column of Figure 8, we can observe that the structure that was an axon in the early time point has been deformed and detached from the cell body. Such sub-cellular dynamics can be observed using our method with time-lapse imaging.

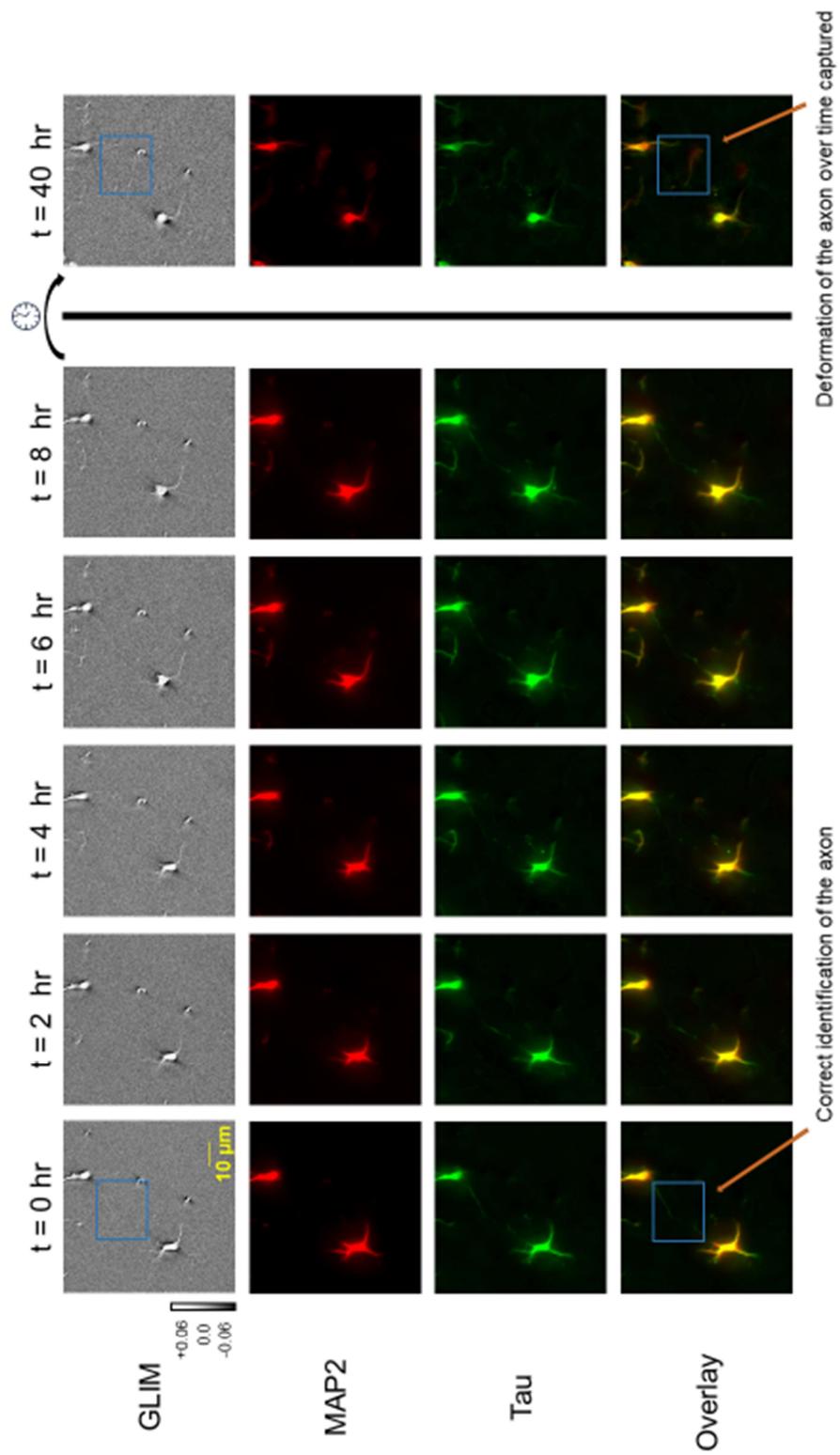

**Figure 8. A time-lapse sequence of the computationally generated fluorescent labels on unseen images.**

## 4.2 Semantic Segmentation

The inference result as shown in the previous section can be used to generate a semantic segmentation map for the neuronal sub-cellular compartments. The semantic segmentation map will classify each pixel of the image into the given categories and assign corresponding values. For the purpose of the semantic segmentation and the neuronal growth analysis, we added another stain called DAPI that labels the nucleus of the cell. Figure 9(a) shows the example of generating a semantic segmentation map using our method from the time-lapse image data. The semantic segmentation was performed using several filter responses and thresholding methods. Figure 9(b) illustrates the labels of the semantic segmentation map and how they are calculated using the computationally generated stain labels. For example, the label for axon is obtained by the area that is labeled by Tau but not by MAP2.

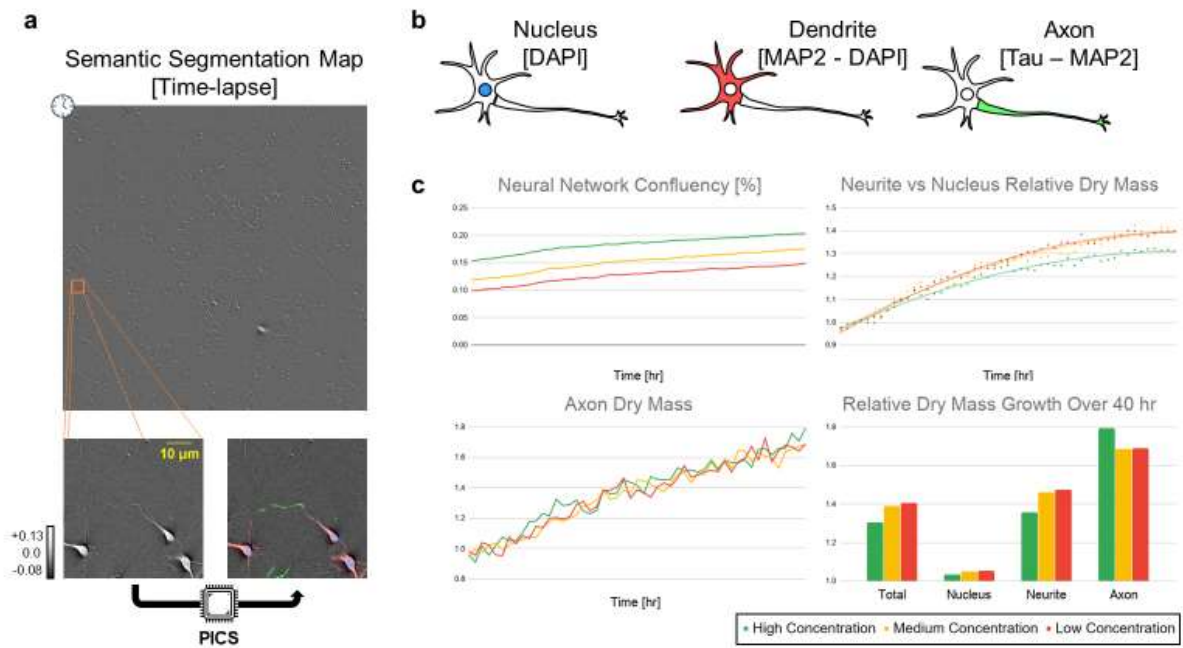

Figure 9. Applications. (a) A semantic segmentation performed on time-lapse data. (b) Illustrations of labels in the semantic segmentation map and definitions. (c) Plots of dry mass analysis on the neuron cultures.

## 4.3 Neuronal Growth Analysis

The semantic segmentation map generated on the time-lapse phase images was used to yield the cellular confluence and dry mass. It enables us to track the cellular confluence and dry mass of the neuron culture over 40 hours. Neuron cultures are divided into three wells or fields of view with varying concentration and the colors in the figure represents the fields of view. The dry mass was normalized by the average measurement from the first 5 hours. Figure 9(c) shows the change in the cell confluency and dry mass of sub-cellular compartments, such as axons and nucleus, over time. We observe that the axons or neurites that includes both axons and dendrites grow much faster, while the nuclei don't grow as much over time. We can also find the neurite versus nucleus relative growth is affected by the concentration where the neurite growth seems inhibited in the higher concentration. The result also hints that there could be the maximum level of neurite growth potential conditioned upon the nucleus dry mass since the neurite versus nucleus growth shows the asymptotic behavior.

# 5. Conclusion

The results reported in this thesis suggest that phase imaging with computational specificity is a promising approach for the time-lapse imaging of neurons. The study of the neural network growth and proliferation requires a high level of specificity that allows the researcher to identify the particular sub-cellular structures of interest in the neural network population such as axons and dendrites. The conventional immunofluorescence techniques suffered the limitations that cells could be damaged during the staining process and it is hard to capture the cellular behavior with fluorescence imaging over the long time period. The result of the time-lapse inference demonstrates that phase imaging with computational specificity can circumvent such problems by generating the fluorescent labels solely using computational methods.